\documentclass[twocolumn, secnumarabic, amssymb, amsmath,nobibnotes, aps, prl,nofootinbib]{revtex4}

\usepackage{graphicx}

\date{\today}

\newcommand{\half}{\mbox{\small{$\frac{1}{2}$}}}
\newcommand{\fourth}{\mbox{\small{$\frac{1}{4}$}}}

\newcommand{\baru}{\bar{u}}
\newcommand{\barv}{\bar{v}}
\newcommand{\barw}{\bar{w}}
\newcommand{\barz}{\bar{z}}
\newcommand{\barr}{\bar{r}}

\begin{document}
\title{the relativity of hyperbolic space}
\author{B. H. Lavenda}
\email{bernard.lavenda@unicam.it}
\affiliation{Universit$\grave{a}$ degli Studi, Camerino 62032 (MC) Italy}
\begin{abstract}
The longitudinal Doppler shift is a measure of hyperbolic distance. Transformations of uniform motion are determined by the Doppler shift, while its square root transforms to a uniformly accelerated frame.  A time-velocity space metric is derived, by magnifying the Beltrami coordinates with the geometric time, which is similar to the one obtained by Friedmann using Einstein's equations in which the mass tensor describes a universe of dust at zero pressure. No such assumption nor any approximation in which the coordinates increase with time (i.e., constant velocities) need be made. The hyperbolic velocities are related to the sides of a Lambert quadrilateral. In the limit when the acute angle becomes an ideal point, the case of uniform acceleration arises. The relations to Hubble's law, and to the exponential red shift, are discussed. \end{abstract}
 
\maketitle

\noindent
\lq\lq Of course, since Einstein, we do not use hyperbolic geometry to model the geometry of the universe\rq\rq \cite{greenberg}.

\section{hyperbolic geometry and the doppler effect}

Projective geometry arose from the need to create three dimensional images in two dimensions.  It differs from Euclidean geometry in two fundamental aspects: the existence of \lq ideal points\rq\ where parallels meet, and transformations which change both length and angle. The notion of a ideal point  can be traced back to Kepler in which an ideal point on each line closes the line into a circle of infinite radius, now called a horocycle. In a perspective drawing all lines in a family of parallels have the same ideal point, known as the horizon.

The question arose as to what remains invariant in a projection since lengths and angles do not. Since three points on a line are not invariant, because it is possible to project them on to another line, the minimum number points needed is four. But then any other four points projectively related to the original ones will have the same cross-ratio since a projectivity is the product of a sequence of perspectivities. The invariant cross-ratio is related to the hyperbolic length through its logarithm. We will evaluate the hyperbolic measure of length in terms of its Euclidean measure through the definition of the cross-ratio.

Ideal points  lie on a circle of radius $1$. Hyperbolic motions are projectivities which are Euclidean rotations about the center of the circle~\cite{busemann}. The circle whose center is $o$ has a Euclidean radius $\overline{r}<1$, and this is also a hyperbolic circle with the same center but a different radius $r$. The problem is to express the Euclidean measure of distance $\overline{r}$ in terms of the hyperbolic measure $r$. Let $\baru<1$ be the Euclidean distance from the origin lying on the same line as the ideal points, $a$ and $b$, located on diametrically opposite ends of the circle as shown in figure 1. 
If $e(x,y)$ denotes the Euclidean distance between points $x$ and $y$, the cross-ratio is:
\begin{eqnarray}
R(o,\baru,b,a)& = & \frac{e(o,b)e(\baru,a)}{e(o,a)e(\baru,b)}\nonumber\\
& = & \frac{1\cdot(1+\baru)}{1\cdot(1-\baru)}. \label{eq:x-ratio}
\end{eqnarray}

The cross-ratio \eqref{eq:x-ratio} is related to the \lq distance\rq\ $u$ in hyperbolic space according to
\begin{equation}
u=\half\ln\left(\frac{1+\baru}{1-\baru}\right), \label{eq:hyper}
\end{equation}
where the scaling constant is included in the nondimensional velocity, $\baru<1$. As the velocity of light tends to infinity, or what is the same $\baru\ll 1$, hyperbolic geometry transforms into Euclidean geometry.

Exponentiating both sides of \eqref{eq:hyper} relates the exponential Poincar\'e distance to the radial Doppler factor
\begin{equation}
e^{u}=\left(\frac{1+\baru}{1-\baru} \right)^{1/2}=:K. \label{eq:Doppler}
\end{equation}
This implies that we should expect a velocity space metric in the case where the Euclidean measure of the velocity, $\baru$, is not constant. Time will act as a magnification factor for the independent velocities rather than as a completely independent coordinate in space-time as in the \lq general\rq\ theory of relativity.
\section{geometry of Doppler and aberration phenomena}

If $\baru$ represents the ratio of the velocity of an object to that of light, then it becomes apparent from \eqref{eq:hyper} that light has a finite speed only because we are using a Euclidean measure of it. This restriction disappears when we use a hyperbolic measure of it, as in \eqref{eq:hyper}. Solving for the Euclidean measure of the velocity in terms of its hyperbolic measure, we obtain
\begin{equation}
\baru=\tanh u. \label{eq:u}
\end{equation}
Consider the triangle formed by rotating $\baru$ through an angle $\vartheta$, as shown in  figure 1. Rotations about the origin do not cause deformations, and there is no difference between Euclidean and hyperbolic measure of the angle. The right triangle has a hypotenuse $\delta$ and height $\alpha$. The cosine of the angle is the same in both measures
\begin{equation}
\cos\vartheta=\cos\overline{\vartheta}=\baru/\overline{\delta}=\tanh u/\tanh\delta. \label{eq:cos}
\end{equation}
However, the opposite angle, $\varphi$, will undergo a contraction  so that it will only be true that
\begin{equation}
\cos\varphi=\overline{\alpha}/\overline{\delta}=\tanh\alpha/\tanh\delta. \label{eq:cos-bis}
\end{equation}

In order to determine the relation between $\varphi$ and $\overline{\varphi}$, it is necessary to calculate the height $\alpha$. If $w$ and $z$ are the corresponding ideal points by extending the height, $e(a,b)$, so that it cuts the circle then the cross-ratio is:
\begin{eqnarray}
R(a,b,w,z) & = & \frac{e(a,w)e(b,z)}{e(a,z)e(b,w)}\nonumber\\
& = & \frac{\sqrt{1-\baru^{2}}\cdot\left(\sqrt{1-\baru^{2}}+\overline{\alpha}\right)}{\sqrt{1-\baru^{2}}\cdot\left(\sqrt{1-\baru^{2}}-\overline{\alpha}\right)}. \label{eq:x-ratio-bis}
\end{eqnarray}
We thus find the Euclidean height in terms of the hyperbolic measure of height as
\begin{equation}
\overline{\alpha}=\gamma^{-1}\tanh\alpha. \label{eq:alpha}
\end{equation}
If the rotation occurred about the origin then $\overline{\alpha}$ would have been $\tanh\alpha$. But because the motion is not at the origin, the Euclidean length will appear contracted by a factor $\gamma^{-1}$. It is precisely this contraction which is responsible for the triangle defect in hyperbolic geometry since
\[
\cos\overline{\varphi}=\overline{\alpha}/\overline{\delta}=\gamma^{-1}\frac{\tanh\alpha}{\tanh\delta}=\gamma^{-1}\cos\varphi.\]
Since the cosine is a decreasing function on the open interval $(0,\pi)$, it follows that $\varphi<\overline{\varphi}$, and this is the origin of the triangle defect in hyperbolic triangles. It is caused by the motion perpendicular to the direction of motion. This is the origin of the Lorentz contraction in the direction perpendicular to the motion~\cite{lavenda}.

In hyperbolic geometry, the Pythagorean theorem, $\overline{\delta}^2=\overline{\alpha}^2+\baru^2$, is converted into
\begin{equation}
\cosh\delta=\cosh\alpha\,\cosh u \label{eq:pythag}
\end{equation}
because
\begin{subequations}
\begin{align}
\tanh^2\delta=\tanh^2\alpha+\tanh^2u \nonumber\\
\mbox{sech}^2\delta=\mbox{sech}^2\alpha+\mbox{sech}^2u, \nonumber
\end{align}
\end{subequations}
and both $\mbox{sech}$ and $\cosh$ are both positive functions. The hyperbolic Pythagorean theorem can be used in
\[\sin\vartheta=\overline{\alpha}/\overline{\delta}=\gamma^{-1}\tanh\alpha/\tanh\delta,\]
to get
\begin{equation}
\sin\vartheta=\frac{\sinh\alpha}{\sinh\delta}. \label{eq:sin}
\end{equation}

Let us now consider what happens in the limits $\alpha,\delta\rightarrow\infty$, such that their difference $\delta-\alpha$ is a positive constant. In this limit,
\[\overline{\alpha}=\gamma^{-1}=1/\cosh\beta=\sin\vartheta=e^{\alpha-\delta}.\]

We can consider a more general triangle with sides $\overline{\alpha}$ and $\overline{\delta}$, and base $\baru$. The altitude $\overline{h}$ cuts the base into two parts $\overline{\varepsilon}$ and $\baru-\overline{\varepsilon}$. The angles formed from the sides and the base are $\vartheta$ and $\varphi$. The sines of these angles are $\sin\vartheta=\overline{h}/\overline{\alpha}=\tanh h\;\mbox{sech}\varepsilon/\tanh\alpha=\sinh h/\sinh\alpha$, and $\sin\varphi=\overline{h}/\overline{\delta}=\tanh h\;\mbox{sech}(u-\varepsilon)/\tanh\delta=\sinh h/\sinh\delta$, since deformation only occurs normal to the direction of motion, i.e., $\baru$. Introducing the Pythagorean theorem of the first triangle, 
\[\cosh\alpha=\cosh\varepsilon\,\cosh h\] into the Pythagorean theorem for the second triangle,
\begin{eqnarray*}
\cosh\delta & = & \cosh h\cosh(u-\varepsilon)\\
& = & \cosh h\{\cosh u/\cosh\varepsilon\\
&- &\sinh\varepsilon\sinh u\},\end{eqnarray*}
results in
\begin{eqnarray*}
\lefteqn{\cosh\delta=\cosh\alpha/\cosh u}\\
& & -\tanh\varepsilon\sinh u/\cosh\alpha.\end{eqnarray*}
Finally, introducing $\cos\vartheta=\tanh\varepsilon/\tanh\alpha$ results in the law of cosines
\begin{eqnarray}
\lefteqn{\cosh\delta=}\label{eq:loc-1}\\
& & \cosh u\cosh\alpha-\sinh\alpha\sinh  u\cos\vartheta. \nonumber
\end{eqnarray}
In an exactly analogous way we find 
\begin{eqnarray}
\lefteqn{\cosh\alpha=}\label{eq:loc-2}\\
& & \cosh\delta\cosh\beta-\sinh\delta\sinh\beta\cos\varphi. \nonumber
\end{eqnarray}

Now, introducing $\cos\vartheta=\tanh\varepsilon/\tanh\alpha$ into $\cos\varphi=\tanh(u-\varepsilon)/\tanh\delta$ results in:
\begin{equation}
\tanh\delta/\cos\varphi=\frac{\tanh u-\tanh\alpha/\cos\vartheta}{1-\tanh u\tanh\alpha\cos\vartheta}.\label{eq:v-comp}
\end{equation}
But, this should be a velocity composition law, and it will become one when we introduce the velocity components $\baru_1=\overline{\alpha}=\tanh\alpha$, and $\baru_2=\overline{\delta}=\tanh\delta$. Introducing these definitions into \eqref{eq:v-comp} gives
\begin{equation}
\baru_2\cos\varphi=\frac{\baru-\baru_1\cos\vartheta}{1-\baru\baru_1\cos\vartheta}. \label{eq:v-comp-bis}
\end{equation}
In the limit as $\alpha,\gamma\rightarrow\infty$, $\baru_1,\baru_2\rightarrow1$, and they become light signals.

The hyperbolic cosine law, \eqref{eq:loc-1} can be written as:
\begin{eqnarray}
\lefteqn{\frac{\sinh\delta}{\sinh\alpha}=}\label{eq:aber}\\
& & \frac{\tanh\delta}{\tanh\alpha}\cosh u\left(1-\tanh\alpha\tanh\beta\cos\vartheta\right)
= \frac{\sin\vartheta}{\sin\varphi}, \nonumber
\end{eqnarray}
which is the aberration formula.
Taking the differential of \eqref{eq:v-comp-bis}, 
\begin{equation}
-\sin\varphi\,d\varphi=\frac{\baru_1}{\baru_2}\frac{\gamma^{-2}\sin\vartheta}{\left(1-\baru\baru_1\cos\vartheta\right)^2}\,d\vartheta, \label{eq:diff}
\end{equation}
and introducing \eqref{eq:aber} result in
\begin{equation}
d\varphi=-\frac{\sqrt{1-\baru^{2}}}{1-\baru\tanh\varepsilon}d\vartheta, \label{eq:diff-bis}
\end{equation}
where we used $\tanh\alpha\cos\vartheta=\tanh\varepsilon$. Dividing both sides by the time increment gives the Doppler shift as
\begin{equation}
\nu=\mathfrak{D}\nu_0, \label{eq:doppler}
\end{equation}
where
\begin{equation}
\mathfrak{D}=\frac{\sqrt{1-\baru^2}}{1-\baru\baru_1\cos\vartheta} \label{eq:D}
\end{equation}
is the Doppler factor. A moving object that emits a signal at frequency $\nu_0=d\vartheta/dt$ with velocity $\baru_1$, and $\nu=-d\varphi/dt$ is the frequency with the observer at rests registers. If the signal is emitted at the velocity of light, $\baru_1=1$, implying that $\alpha\rightarrow\infty$, and $\vartheta\rightarrow\pi/2$, or equivalently $\varepsilon\rightarrow0$,  then it follows from \eqref{eq:loc-1} that $\delta\rightarrow\infty$ such that the difference $\delta-\alpha$ remains finite, i.e.,
\begin{equation}
e^{(\delta-\alpha)}=\cosh u. \label{eq:exp}
\end{equation}
The Doppler shift \eqref{eq:doppler} then becomes the \emph{exponential\/} Doppler shift
\begin{equation}
\nu=e^{-(\delta-\alpha)}\nu_0. \label{eq:doppler-exp}
\end{equation}

Ordinarily, one writes the Doppler factor \eqref{eq:D} with $\baru_1=1$ without realizing that it requires the limit $\alpha\rightarrow\infty$, which, in turn, requires that it be perpendicular to the motion. Moreover, \eqref{eq:exp} is the well-known expression for angle of parallelism: the ratio of concentric limiting arcs between two radii is the exponential distance between the arcs divided by the radius of curvature.

\section{kinematics: $K$-calculus}
With the realization that there is no such thing as a rigid body in relativity, Whithrow~\cite{gjw}  went on to develop a radar method, or what he called a \lq signal-function method\rq\  where light signals are transmitted between different inertial frames, and non-inertial ones. It was afterwards referred to as the \lq $K$-calculus technique\rq\ by Bondi~\cite{bondi}, although Milne~\cite{milne} used it extensively in his research prior to him.
\subsection{constant relative velocity: geometric-arithmetic mean equality}
As in kinematical relativity~\cite{milne}, time measurements are much more fundamental than distance measurements, the latter being deducible from the former. In other words, distances are measured by the elapse of time. This has been criticized by Born~\cite{born} as being impractical since no one has ever received light signals from nebulae beyond the horizon. However, it is far superior to the usual method in general relativity that uses a metric, or rigid ruler,  to measure distance~\cite{bondi}. What was discarded in \lq special\rq\ relativity  made its come back in \lq general\rq\ relativity. 

The most ideal situation would be to introduce into the fabric of the theory distances measured in brightness, or the difference between apparent and absolute brightness. However, no one has succeeded in doing so and we will base all distance measurements on the so-called radar method~\cite{gjw}, where a light signal is sent out and reflected at a later time. All that is needed is that at each reflection a certain factor $K$ comes in which is determined by the clock in the frame that is sending out the light pulse.

For consider two observers, $A$ and $B$. Observer $A$ sends out a light signal in his time $t^{A}_{1}$ which is received by observer $B$ in his time $t^{B}_{2}$. In terms of $A$'s time, $B$ will receive it in $Kt^{A}_{1}$, where $K$ is some constant factor that is a function only of the relative velocity of the two frames. The signal that passes $B$ in time $t^{B}_{2}$ will be reflected at some later time. The reflected signal passes $B$ in time $t^{B}_{3}$ which arrives at observer $A$ in time $t^{A}_{4}$, where $t^{A}_{4}=Kt^{B}_{3}$. From this it is apparent that both observers will call the reflection time
\begin{equation}
t^{r}=\sqrt{t^{A}_{1}t^{A}_{4}}=\sqrt{t^{B}_{2}t^{B}_{3}}, \label{eq:geo}
\end{equation}
which is the geometric mean of the time intervals, and it is an invariant independent of the frame. So the \lq signal-function method\rq\ of Whithrow singles out the geometric mean as the time of reflection.

The reading shown by a synchronous (stationary) clock at the event should be midway between the observer's time, $t^{A}_{1}$, of sending out the signal, and the time he receives its reflection, $t_{4}^{A}$, viz.,
\begin{equation}
t=\half\left(t^{A}_{1}+t^{A}_{4}\right). \label{eq:t}
\end{equation}
This was Einstein's choice, but it by no means is the only choice~\cite[\S 5.2]{whithrow}.
The measure of the space interval is the difference between  the \lq\lq average\rq\rq\ for the light-signalling process, \eqref{eq:t}, and the time the signal was sent out
\begin{equation}
r=t-t_{1}^{A}=\half\left(t^{A}_{4}-t^{A}_{1}\right). \label{eq:r}
\end{equation}

In terms of $B$'s coordinates, he will measure a time interval
\begin{equation}
t^{\prime}=\half\left(t_{2}^{B}+t_{3}^{B}\right), \label{eq:t-prime}
\end{equation}
and a space interval 
\begin{equation}
r^{\prime}=\half\left(t_{3}^{B}-t_{2}^{B}\right), \label{eq:r-prime}
\end{equation}
separating the event  from where he is located. The two systems of inertial coordinates $(t,r)$ and $(t^{\prime},r^{\prime})$ are related by: 
\begin{subequations}
\begin{align}
t_{2}^{B}=t^{\prime}-r^{\prime}=K(t-r)=Kt_{1}^{A}\label{eq:L1}\\
t_{3}^{B}=t^{\prime}+r^{\prime}=K^{-1}(t+r)=K^{-1}t_{4}^{A}.\label{eq:L2}
\end{align}
\end{subequations}
The time $t_{2}^{B}$ is the time on $B$'s clock when the signal is received, and $t_{3}^{B}$ is the moment on $B$'s clock when it is sent back.  

Suppose, for the  moment, we place ourselves at the origin of $B$'s frame. Then summing \eqref{eq:L1} and \eqref{eq:L2} gives
\begin{equation} 
t=\half\left(t_{1}^{A}+t_{4}^{A}\right)=\half\left(K+K^{-1}\right)t^{\prime}=\frac{t^{\prime}}{\sqrt{1-\baru^{2}}} ,\label{eq:t-di}
\end{equation}
showing that a clock traveling at a uniform velocity goes slower than one at rest. This expression for time-dilatation only holds for frames moving at a constant velocity $\baru$ [cf. eqn \eqref{eq:t-di-bis} below].

In terms of the longitudinal Doppler shift, \eqref{eq:Doppler}, the two system of coordinates are related by:
\begin{eqnarray}
t &=& \half\left(t_{1}^{A}+t_{4}^{A}\right)=\half\left(Kt_{3}^{B}+K^{-1}t_{2}^{B}\right)\nonumber\\
&=& \half\left\{(K+K^{-1})t^{\prime}+(K-K^{-1})r^{\prime}\right\}\nonumber\\
&=& t^{\prime}\cosh u+r^{\prime}\sinh u, \label{eq:L-t}
\end{eqnarray}
and
\begin{eqnarray}
r &=& \half\left(t_{4}^{A}+t_{1}^{A}\right)=\half\left(Kt_{3}^{B}-K^{-1}t_{2}^{B}\right)\nonumber\\
&=& \half\left\{(K-K^{-1})t^{\prime}+(K+K^{-1})r^{\prime}\right\}\nonumber\\
&=& t^{\prime}\sinh u+r^{\prime}\cosh u, \label{eq:L-r}
\end{eqnarray}
These are none other than the well-known Lorentz transformations. Taking their differentials and forming the difference of their squares shows that the hyperbolic form
\begin{equation}
dt^{2}-dr^{2}=dt^{\prime\,2}-dr^{\prime\, 2} \label{eq:L-metric}
\end{equation}
is invariant.

Now let us ask what happens when the light-signal is reflected when it arrives at $B$. In this case, $t_{2}^{B}=t_{3}^{B}\equiv t^{r}$ is the time of reflection, and it occurs at the same point in space for $B$ so that $r^{\prime}=0$. The Lorentz transformations, \eqref{eq:L-t} and \eqref{eq:L-r}, degenerate into:
\begin{subequations}
\begin{align}
t=t^{r}\cosh u \label{eq:l-t}\\
r=t^{r}\sinh u \label{eq:l-r}.
\end{align}
\end{subequations}
Equation \eqref{eq:l-t} is a statement of the arithmetic-geometric mean inequality: the arithmetic mean $t$ can never be inferior to the geometric mean $t^{r}$ since $\cosh u\ge1$.
Adding and subtracting the equations \eqref{eq:l-t} and \eqref{eq:l-r} give:
\begin{subequations}
\begin{align}
t+r=Kt^{r} \label{eq:t+r}\\
t-r=K^{-1}t^{r}.\label{eq:t-r}
\end{align}
\end{subequations}
Taking the differentials of \eqref{eq:t+r} and \eqref{eq:t-r} and then the product of the two, without requiring that $K$ be constant, result in:
\begin{equation}
dt^{2}-dr^{2}=dt^{r\,2}-t^{r\,2}du^{2}. \label{eq:l-metric}
\end{equation}
A space-time interval has been transformed into a velocity space-time interval.

\subsection{constant relative acceleration}

It is generally acknowledged that acceleration has no effect on the rate of a clock \cite{moller}, and that the expression for time-dilatation \eqref{eq:t-di} can be used in its infinitesimal form  whether or not $\baru$ is constant. However, according to Einstein's equivalence principle uniform acceleration is equivalent to, or indistinguishable from, a uniform gravitational field. It has been shown from the gravitational red-shift that the latter, indeed, has an effect on the rate of a clock. This contradiction has been clearly pointed out by Whithrow \cite{whithrow}, who shows that the time-dilatation is greater when the velocity is varying with time than when it is constant. We convert his inequality into an equality.
   
For a particle under the influence of a constant gravitational acceleration,
\begin{equation}
g=\frac{d}{dt}\left(\frac{\baru}{\sqrt{1-\baru^{2}}}\right), \label{eq:g}
\end{equation}
will be constant so that integration gives simply:
\begin{equation}
gt=\frac{\baru}{\sqrt{1-\baru^{2}}}=\sinh u. \label{eq:sinh}
\end{equation}
Now, the velocity can be written as:
\begin{equation}
\baru=\tanh u=\frac{gt}{\sqrt{1+(gt)^{2}}}=\frac{dr}{dt}. \label{eq:tanh}
\end{equation}
If we further assume that $r=0$ at $t=0$, we get a second integral as:
\begin{equation}
gr=\sqrt{1+(gt)^{2}}-1=\cosh u-1. \label{eq:cosh}
\end{equation}
This is the one-dimensional hyperbolic motion found by Born in 1909, and by Sommerfeld a year later~\cite[\S 29]{moller}. This will be our prototype of one-dimensional systems at constant acceleration.

Dividing \eqref{eq:sinh} by \eqref{eq:cosh} leads to the average velocity
\begin{equation}
r/t=\tanh(u/2). \label{eq:r/t}
\end{equation}
Consider two observers receding from one another with an average velocity $r/t$. Their identical clocks were synchronized at $t^A=t^B=0$ when they were at the same point. At time $t^A_1$, A emits a signal which is picked up and immediately reflected by B at time $t^{B\,r}$, and received back at A at time $t^A_3$. The space interval is 
\[t-t_1^A=t^A_3-t=\half\left(t^A_3-t^A_1\right)=r.\]
From this it follows that
\begin{subequations}
\begin{align}
t^A_1=t-r\label{eq:t1}\\
t^A_3=t+r\label{eq:t3}\\
t^{A\,r}=\sqrt{1-(r/t)^2}\;t. \label{eq:tr}
\end{align}
\end{subequations}

Since the Doppler shift is
\begin{equation}
K=\left(\frac{1+\baru}{1-\baru}\right)^{1/2}=\left(\frac{1+r/t}{1-r/t}\right), \label{eq:K-bis}
\end{equation}
we can express \eqref{eq:t1} and \eqref{eq:t3} as $t^A_3=Kt^A_1$, or
\begin{subequations}
\begin{align}
t^{B\,r}=K^{1/2}t_{1}^A ,\label{eq:K1}\\
t_3^A=K^{1/2}t^{B\,r}. \label{eq:K3}
\end{align}
\end{subequations}
But, from \eqref{eq:tr} it is apparent that $t^{B\,r}=t^{A\,r}$ so that the clocks remain synchronized, and we can  drop the superscripts on the time.

Expressing $r$ and $t$ in terms of $t_{1}$ and $t_{3} $ in  \eqref{eq:cosh} we find~\cite{page}:
\begin{equation}
g=2\frac{r}{t^{r\,2}}=\frac{1}{t_1}-\frac{1}{t_3}. \label{eq:page}
\end{equation}
Employing \eqref{eq:K1} and \eqref{eq:K3} we write \eqref{eq:page} as:
\begin{equation}
g=\frac{K^{1/2}-K^{-1/2}}{t^r}=\frac{2\sinh(u/2)\cosh(u/2)}{t}, \label{eq:sinh-bis}
\end{equation}
which is identical to \eqref{eq:sinh}. 

Equation \eqref{eq:page} enables us to express the Doppler shift, $K$, in terms of the ratio of the time the signal was received back to that when it was sent out, viz.,
\begin{equation}
t_3/t_1=K. \label{eq:tK}
\end{equation}
Taking the logarithms of both sides of \eqref{eq:tK} and then differentiating with respect to $t$, give:
\[
\frac{d\ln t_3}{dt}-\frac{d\ln t_1}{dt}=\frac{1+\baru}{t_3}-\frac{1-\baru}{t_1}=\frac{1}{1-\baru^2}\frac{d\baru}{dt},\]
where we have used the differentials of \eqref{eq:t1} and \eqref{eq:t3}. Dividing both sides by $\sqrt{1-\baru^2}$ results in
\begin{equation}\frac{K}{t_3}-\frac{K^{-1}}{t_1}=\frac{1}{\sqrt[3]{1-\baru^2}}\frac{d\baru}{dt}=g,\label{eq:g-bis}
\end{equation}
which is identical to \eqref{eq:page}. If the ratio \eqref{eq:tK} had been proportional to the square of the Doppler shift,  we would have found that \eqref{eq:g-bis} vanishes.

If we consider  $t^{\prime}$ to be the time of reflection on $B$'s clock, we can write \eqref{eq:page} as:
\begin{subequations}
\begin{align}
\frac{1}{t_{1}}=\frac{1}{t^{\prime}}+g/2 \label{eq:page1}\\
\frac{1}{t_{3}}=\frac{1}{t^{\prime}}-g/2. \label{eq:page2}
\end{align}
\end{subequations}
Adding \eqref{eq:page1} and \eqref{eq:page2} shows that \emph{the time of reflection on $B$'s clock is the harmonic mean\/},
\begin{equation}
\frac{1}{t^{\prime}}=\half\left(\frac{1}{t_{1}}+\frac{1}{t_{3}}\right), \label{eq:harmonic}
\end{equation}
\emph{in contrast with the geometric mean as the time of reflection for uniform motion\/}. The difference between \eqref{eq:page1} and \eqref{eq:page2} is \eqref{eq:page}. Writing $t_{1}=t-r$ and $t_{3}=t+r$ in     
 \eqref{eq:harmonic} clearly shows that the space-time interval is not constant,
\[t^{2}-r^{2}=t^{r\;2}=t^{\prime}t=t^{r\;2},\]
unless we require the reflection times to be the same, meaning that clocks $A$ and $B$ are synchronous \cite{pro}.

Multiplying the left- and right-hand sides of \eqref{eq:page1} and \eqref{eq:page2} , rearranging, and then taking the square roots give
\begin{equation}
t^{\prime}=\mbox{sech}(u/2) t^{r}=\mbox{sech}^{2}(u/2) t, \label{eq:har-geo}
\end{equation}
where the second equality follows from \eqref{eq:tr}.  Equations \eqref{eq:har-geo} give quantitative relationships to the statements that the harmonic mean is always smaller than the geometric mean which is smaller than the arithmetic mean, because the equality of times can never apply. The first equality in equation \eqref{eq:har-geo} states physically that the time of reflection on $B$'s clock is always less than on $A$'s clock.

From \eqref{eq:page1} and \eqref{eq:page2}  it also follows that
\begin{equation}
K=\frac{t_{3}}{t_{1}}=\frac{1+gt^{\prime}/2}{1-gt^{\prime}/2}. \label{eq:K-tris}
\end{equation}
A comparison of \eqref{eq:K-bis} and \eqref{eq:K-tris} leads to
\[
\half gt^{\prime}=r/t. \]
Differentiating with respect to the arithmetic time average gives:
\begin{equation}
dt=\cosh^{2}(u/2)\frac{dt^{\prime}}{\sqrt{1-\baru^{2}}}. \label{eq:t-di-bis}
\end{equation}
In comparison with the expression for time-dilatation for uniform motion, \eqref{eq:t-di}, expression \eqref{eq:t-di-bis} is larger for uniform acceleration. Uniform acceleration of $B$ does, indeed, affect the apparent rate according to $A$ of a clock carried by $B$ \cite[pp. 263-4]{whithrow}.
 
The transformation laws \eqref{eq:K1} and \eqref{eq:K3} can be expressed as:
 \begin{subequations}
 \begin{align}
 t+r=K^{1/2}t^{r} \label{eq:l1}\\
 t-r=K^{-1/2}t^{r} .\label{eq:l2}
 \end{align}
 \end{subequations}
  Taking the differentials of \eqref{eq:l1} and \eqref{eq:l2}, and then their product, result in
 \begin{eqnarray}
 ds^{2}:&=&dt^{2}-dr^{2}=dt^{r\;2}-\fourth t^{r\;2}(dK)^{2}/K^{2} \nonumber\\
 &=&dt^{r\;2}-\fourth t^{r\;2}du^{2}.\label{eq:metric}
 \end{eqnarray}
 The appearance of $t^{r\;2}$ in the velocity space component of the metric \eqref{eq:metric} implies uniform expansion. Introducing the logarithmic time
 \begin{equation}
 \tau=2\tau_{0}\ln\left(t^{r}/\tau_{0}\right), \label{eq:tau}
 \end{equation}
 where $\tau_{0}$ is an absolute constant, enables \eqref{eq:metric} to be written as:
 \begin{equation}
 ds^{2}=dt^{2}-dr^{2}=\frac{e^{\tau/\tau_{0}}}{4}\left\{d\tau^{2}-\tau_{0}^{2}du^{2}\right\} .\label{eq:metric-bis}
 \end{equation}
 
 Thus, the formulas of the transformation of coordinates \eqref{eq:l1} and \eqref{eq:l2} can be written as:
 \begin{subequations}
 \begin{align}
 t+r=\tau_{0}e^{\tau/\tau_{0}+u/2} \label{eq:l1-bis}\\
 t-r=\tau_{0}e^{\tau/\tau_{0}-u/2}.\label{eq:l2-bis}
 \end{align}
 \end{subequations}
 For constant $t^{r}$, the surface of the equation is obtained by rotating the hyperbola $t^{2}-r^{2}=t^{r\;2}$ around the $t$ axis to give a bowl shaped form. 

The equivalence relations \eqref{eq:l1-bis} and \eqref{eq:l2-bis} can be combined to read:
\begin{equation}
t+r=K(t-r). \label{eq:l-acc}
\end{equation}
Comparing this with the case of constant velocity,
\begin{subequations}
\begin{align}
t+r=K\left(t^{\prime}-r^{\prime}\right) \label{eq:l1-vel}\\
t-r=K^{-1}\left(t^{\prime}+r^{\prime}\right), \label{eq:l2-vel}
\end{align}
\end{subequations}
we conclude that the former does  not retain its invariant hyperbolic form whereas the latter does,
\begin{equation}
t^2-r^2=t^{\prime\;2}-r^{\prime\;2}. \label{eq:h}
\end{equation}
 Adding and subtracting the equations yields the well-known Lorentz transformations \eqref{eq:L-t} and \eqref{eq:L-r}, and from which it can be concluded that the Lorentz transformations leave invariant the hyperbolic \lq distance\rq, \eqref{eq:h}.

In terms of radar measurements, \eqref{eq:l-acc} consists in a single observer: a light pulse is emitted in time $t_1$, and observed by him at a later time $t_2=Kt_1$. Alternatively, in the case of constant velocity, \eqref{eq:l1-vel} says a light signal is emitted at time $t^{\prime}_1$, in the prime inertial frame, and observed in the unprimed frame at a later time $t_2=Kt_1^{\prime}$. Whereas, \eqref{eq:l2-vel} says that if a signal is emitted at time $t_1$, it will be observed at time $t_2^{\prime}$ in the primed inertial frame. 

For uniform motion the geometric mean times, $\sqrt{t_1t_2}=\sqrt{t_1^{\prime}t_2^{\prime}}$ remain invariant. This is the same as requiring the hyperbolic line element \eqref{eq:h} be invariant.  While, for uniform acceleration, the time of reflection in the $B$ frame is the harmonic mean of the $A$ frame.  In his analysis of uniform acceleration, Page~\cite{page} attempted to show that the space-time interval between neighboring points is not constant. His analysis replaces \eqref{eq:page2} by
\begin{equation}
\frac{1}{t_{3}}=\frac{1}{t^{\prime\prime}}-g/2. \label{eq:page2-bis}
\end{equation}
This condition would necessarily imply that the harmonic means in the two frames are equal. Solving \eqref{eq:page1} and \eqref{eq:page2-bis} for the times $t^{\prime}$ and $t^{\prime\prime}$, with $t^{\prime\prime}>t^{\prime}$ we get:
\begin{subequations}
\begin{align}
\tilde{t}:=\half\left(t^{\prime}+t^{\prime\prime}\right)=t\;\mbox{sech}^{2}(u/2) \label{eq:lav1}\\
\tilde{r}:=\half\left(t^{\prime\prime}-t^{\prime}\right)=\left[r-t\tanh(u/2)\right]\mbox{sech}^{2}(u/2). \label{eq:lav2}
\end{align}
\end{subequations}
However, consulting \eqref{eq:r/t}, \eqref{eq:lav2} vanishes, and, hence $\tilde{r}=0$. The time of reflection is given by the harmonic mean \eqref{eq:harmonic}. Therefore, for uniformly accelerating systems the point of reflection must occur at the origin of  $B$'s frame, whose time is given by the harmonic mean of $A$'s clock.

\section{Multi-dimensional Lobachevsky velocity space}

The one-dimensional velocity space of the previous subsection can be generalized to at least two velocities by realizing that \eqref{eq:tanh} is a Beltrami coordinate for the velocity. A second Beltrami coordinate can be defined similarly:
\begin{equation}
\barv=\tanh v. \label{eq:tanh-bis}
\end{equation}
The Weierstrass coordinates can now be introduced as:
\begin{equation}
X=\baru T, \;\;\;\;\; Y=\barv T,\;\;\;\; \mbox{and}\;\;\;\;\; T=\cosh u\cosh w, \label{eq:W}
\end{equation}
where $u$, $v$ $w$, and $z$ are four sides of a Lambert quadrilateral, shown in figure 2, consisting of three right angles and one acute angle between $w$ and $z$. The condition that the two sides will intersect to form an acute angle is
\begin{equation}
1-\tanh^{2}u-\tanh^{2}v=1-\baru^{2}-\barv^{2}>0. \label{eq:ineq}
\end{equation}

The Euclidean measures of the two sides are $\barw=\barv/\sqrt{1-\baru^{2}}$, and $\barz=\baru/\sqrt{1-\barv^{2}}$. By giving to each point the triple $(X,Y,T)$ of Weierstrass coordinates, the hyperbolic plane is mapped on to the locus
\[T^{2}-X^{2}-Y^{2}=1,\]
which is one of two sheets of a hyperboloid in Cartesian three-dimensions. The differential metric
\begin{eqnarray}
\lefteqn{dX^{2}+dY^{2}-dT^{2}=}\nonumber\\
&=& dw^{2}+\cosh^{2}w\,du^{2}\nonumber\\
&=& \frac{(1-\barv^{2})d\baru^{2}+2\baru\barv d\baru\,d\barv+(1-\baru^{2})d\barv^{2}}{(1-\barr^{2})^{2}} ,\label{eq:v-metric}
\end{eqnarray}
where $\barr^{2}=\baru^{2}+\barv^{2}$, is the spatial component of the Lobachevsky velocity space metric. 

In order to derive the full time-velocity space metric, we magnify the $T$ coordinate, $t^{r}>0$ times, viz.,
$T=t^{r}/\sqrt{1-\barr^{2}}$ so as to obtain the time-like, indefinite metric as:
\begin{eqnarray}
ds^{2} &= & dT^{2}-dX^{2}-dY^{2}\nonumber\\
&= & dt^{r\;2}-t^{r\;2}\left(dw^{2}+\cosh^{2}w\,du^{2}\right). \label{eq:v-metric-tris}
\end{eqnarray}

 A time-velocity  metric,  similar to \eqref{eq:metric-bis}, was  derived by Friedmann in 1922 using the Einstein equations to relate the coordinates to the Lagrangian variables, $u_{i}$. It was derived under the condition that matter was \lq\lq dust-like\rq\rq\ exerting zero pressure. It was also assumed that the velocities variables $u_{i}$ are constants relating the spatial coordinates $x_{i}$ to time, but, subsequently, they were differentiated to obtain the Lobachevsky-Friedmann metric \eqref{eq:metric-bis}~\cite{fock}.
 
As can be seen from the definition of the Weierstrass coordinates, \eqref{eq:W}, each of the coordinates become magnified $t^{r}$ times~\cite[eqn (94.47)]{fock}. In a multi-dimensional velocity space, the spatial part of the metric can be written as:
\[d\sigma^{2}=\frac{(d\vec{u})^{2}-(\vec{u}\times d\vec{u})^{2}}{(1-\barr^{2})^{2}},\]
by introducing the coordinates $X_{i}=\baru_{i}t^{r}/\sqrt{1-\barr^{2}}$ into the infinitesimal metric
\[ds^{2}=dT^{2}-\sum_{i}X_{i}^{2}.\]

In relation to the Robertson metric, the scale factor $R(t)$ multiplying the spatial part of the metric is just $t$ meaning uniform expansion. Introducing logarithmic time according to
\begin{equation}
\tau=\tau_{0}\ln\left(t^{r}/\tau_{0}\right), \label{eq:tau-tris}
\end{equation}
(and not \eqref{eq:tau}) into \eqref{eq:v-metric-tris} gives:
\begin{equation}
ds^{2}=e^{2\tau/\tau_{0}}\left\{d\tau^{2}-\tau_{0}^{2}\left(dw^{2}+\cosh^{2}w\,du^{2}\right)\right\} .\label{eq:gen-metric}
\end{equation}
The proper time interval is the quantity $s_{0}$ determined at constant velocity by the equation
\[
s_{0}=\int_{0}^{\tau}e^{\tau/\tau_{0}}d\tau=\tau_{0}\left(e^{\tau/\tau_{0}}-1\right). \]
This law could have been anticipated because $t^{r}$ is the geometric mean. Only for $\tau\ll\tau_{0}$ will the proper time coincide with $\tau$. The exponential variable scale factor multiplies both time and velocity increments, and testifies to the fact that they are not independent, but, are related by the Beltrami coordinates and logarithmic time. 

For fixed $s_{0}$ the velocity space line element is:
\begin{equation}
d\sigma^{2}=\tau_{0}^{2}e^{2\tau/\tau_{0}}\left(dw^{2}+\cosh^{2}w\;du^{2}\right) .\label{eq:sigma}
\end{equation}
The terms in the parentheses have the metric form of a pseudosphere in velocity space, with constant negative curvature, $-1$. The scale factor is the same exponential that appears in the proper time increment. The hallmark of a pseudosphere is that lines which do not intersect are, nevertheless, not parallel.
Along a light track \eqref{eq:gen-metric} vanishes, resulting in
\[d\tau=\tau_{0}
\frac{\sqrt{(1-\baru^{2})^{2}d\barw^{2}+(1-\barw^{2})d\baru^{2}}}{(1-\baru^{2})(1-\barw^{2})}.\]
This is a generalization of the well-known one-dimensional expression, whose integral identifies \eqref{eq:u} as the length of the corresponding segment of a Lobachevsky straight line.

\section{limiting case of a Lambert quadrilateral: uniform acceleration}
A limiting case arises when inequality \eqref{eq:ineq} reduces to an equality:
\begin{equation}\baru^{2}+\barv^{2}=1,\label{eq:equal}
\end{equation}
or $\barv=\sqrt{1-\baru^{2}}=:\baru^{*}$. The velocities $\baru$ and $\baru^{*}$ are said to be complementary~\cite[p. 412]{greenberg}. The defining relation for uniform acceleration is \eqref{eq:K-bis}, which upon resolving for the velocity gives:
\begin{equation}
\baru=\tanh u=\frac{2\ell}{1+\ell^{2}}, \label{eq:ell}
\end{equation}
where, for brevity, we have set $r/t=\ell$. $\ell$ represents the Euclidean length, while $u$ the Poincar\'e length in the Poincar\'e model; the two being related by:
\begin{equation}
e^{u}=\frac{1+\ell}{1-\ell}. \label{eq:exp-bis}
\end{equation}
The complementary velocity is found to be:
\[
\baru^{*}=\frac{1-\ell^{2}}{1+\ell^{2}}=\mbox{sech}u=\sqrt{1-\baru^{2}}, \]
which verifies \eqref{eq:equal}.

 The angle of parallelism is defined as:
 \begin{equation}
 \Pi(u^{*})=2\tan^{-1}e^{-u^{*}}, \label{eq:Pi}
 \end{equation}
 which is defined solely in terms of the \lq distance\rq\ $u^{*}$.
 The angle of parallelism is the lower bound for the angle of parallax.  It was Bernoulli who first showed that 
 \begin{equation}2\tan^{-1}e^{-u^{*}}=\frac{1}{i}\ln\left(\frac{1+ie^{-u^{*}}}{1-ie^{-u^{*}}}\right).
 \label{eq:bernoulli}
 \end{equation}
 In particular,
 \begin{equation}
 r/t=\tanh(u/2)=\tan\left[\Pi(u^{*})/2\right]=e^{-u^{*}}. \label{eq:r/t-tris}
 \end{equation}
 The closer the complementary velocity $u^{*}$ is to zero, the closer $\Pi$ is to being a right angle. For large $u^{*}$, or non relativistic velocities, the angle of parallelism is practically zero. 
 
 Moreover, according to the double angle formula,
 \begin{equation}\tan\Pi(u^{*})=1/\sinh u^{*}=\sinh u,\label{eq:mom}
 \end{equation}
 showing that $\Pi$ provides the link between hyperbolic and circular functions. In particular, \eqref{eq:mom} relates the angle of parallelism to the particle velocity [cf. eqn \eqref{eq:sinh}].
 
 Consider a Lambert quadrilateral with three right angles and an ideal point in figure 3. The opposite right angle is divided into  two angles of parallelism such that
 \begin{equation}
 \Pi(u)+\Pi(u^{*})=\pi/2. \label{eq:Pi-cond}
 \end{equation}
 Using Bernoulli's relation, \eqref{eq:bernoulli},  this implies that the complementary velocities, $u$ and $u^{*}$, which are adjacent to the two angles of parallelism, are related by
 \[
 e^{-u^{*}}=\left(\frac{1-e^{-u}}{1+e^{-u}}\right).\]
 Equation \eqref{eq:exp-bis} implies the addition law for the hyperbolic measure of the complementary velocities implies the product law for the average velocities.  
 
 Rather, if $u_{1}$ and $u_{2}$ are the components of the hyperbolic measure of the velocity $u$, their composition law follows Poincar\'e's addition law
 \begin{equation}
 e^{-u}=\frac{e^{-u_{1}}+e^{-u_{2}}}{1+e^{-u_{1}-u_{2}}} .\label{eq:poincare}
 \end{equation}
 Introducing \eqref{eq:r/t-tris} into \eqref{eq:exp-bis}, and then into \eqref{eq:poincare} lead to the result that
\begin{equation}
\tanh(u/2)=\prod_{i=1}^n\tan\left[\Pi(u_i^{*})/2\right]=e^{-\sum_{i=1}^n u_i^{*}}=\hat{\ell}^n, \label{eq:BL-bis}
\end{equation}
where the Euclidean length $\hat{\ell}$ is the geometric mean of a sample of size $n$. 

We conclude that whereas uniform motion utilizes the arithmetic mean, uniform acceleration calls for the geometric mean. For uniform motion, the geometric mean time is the invariant reflection time, while the harmonic mean time is the reflection time for uniform acceleration.

\section{additivity of the recession and distance in Hubble's law}

The fact that the shift $z=\delta\lambda/\lambda_{0}$ for lines in the spectrum of a given galaxy is independent of the wavelength is a necessary, but not a sufficient condition, that the red-shift is due to motion. It was Hubble who interpreted these red shifts as  Doppler shifts, indicative  of recessional motion. In so doing he obtained a linear relation between the velocity of recession, $u$, and radial distance, $r$, with a constant of proportionality that is the same for all galaxies. We will show that both these quantities are \emph{additive\/}. 

There will be a red-shift if the detected wavelength, $\lambda$, is greater than the emitted wavelength, $\lambda_0$, in
\begin{equation}
1+z=\frac{\lambda}{\lambda_0}=K=e^{u}. \label{eq:z}
\end{equation}
 Now $K$ is the ratio of the received, $t_2$, to the emitted time, $t_1$. We can therefore define a hyperbolic measure of the time interval as~\cite[p. 57]{milne}:
\begin{equation}
\tau/\tau_0=\ln\frac{t_2}{t_1}=\ln K. \label{eq:tau-bis}
\end{equation}
A comparison of \eqref{eq:z} and \eqref{eq:tau-bis} results in
\begin{equation}
u=\frac{\tau}{\tau_0}=H\,r, \label{eq:hubble}
\end{equation}
where $u$ is a hyperbolic measure of the velocity, $H=\tau_0^{-1}$, the Hubble parameter, and $r=\tau$ is the hyperbolic measure of distance.  Hubble's law could have also been obtained by setting the one-dimensional velocity space metric \eqref{eq:metric} equal to zero, and introducing logarithmic time, \eqref{eq:tau-tris}.

Consequently, \eqref{eq:z} is the exponential law~\cite[p. 75]{pro}:
\begin{equation}
1+z=e^{H\tau}. \label{eq:z-bis}
\end{equation}
Only when $H\tau\ll1$ can we neglect  powers of $H\tau$ greater than first so that \eqref{eq:z-bis} reduces to the relation~\cite{fred}:
\begin{equation}
z=H\,\tau. \label{eq:z-linear}
\end{equation}
The exponential law \eqref{eq:z-bis} implies that when there is more than one red-shift, it is their geometric mean which should be taken. For instance, the cluster, Group II has $n=21$ redshifts, in which case \eqref{eq:z} generalizes to:
\begin{eqnarray}
\lefteqn{\prod_{i=1}^n\left(1+z_i\right)=}\nonumber\\
& &\prod_{i=1}^n\frac{\lambda_i}{\lambda_{0i}}=\prod_{i=1}^nK(u_i)=\exp\left(\sum_{i=1}^{n}u_i\right). \label{eq:z-gen}
\end{eqnarray}
The hyperbolic velocities, $u_{i}$, like the hyperbolic distances, $r_{i}$, are additive. The average wavelength is the geometric mean wavelength. This is implied by the exponential law \eqref{eq:z-gen}.

\section{exponential red shift}
Let us again consider signal transmission from observers $A$ and $B$. We know that observer $B$ will receive a signal sent from $A$ in time $t_{2}^{A}=K t_{1}^{A}$. If we further specify that $A$ will receive back the signal in a time interval increased $K$ times again, it means that $B$ reflects the signal upon receiving it because $t_{2}^{B}=t_{3}^{B}$. 

What we want to show now is that the frequency shift, \eqref{eq:doppler}, is the inverse of the ratio of the arithmetic times, and to give a geometrical interpretation of the result. We claim that
\begin{equation}
\frac{\nu}{\nu_0} =  \frac{t_{2}^{B}+t_{3}^{B}}{t_{1}^{A}+t_{4}^{A}}=K\frac{1+K^{-2}\left(t_{4}^{A}/t_{1}^{A}\right)}{1+t_{4}^{A}/t_{1}^{A}}=\mathfrak{D},
 \label{eq:doppler-bis}
\end{equation}
where we are allowing for signals other than light since $\baru_2$ need not be unity. The condition that \eqref{eq:doppler-bis} hold is:
\begin{equation}
\frac{t_{4}^{A}}{t_{1}^{A}}=\left(\frac{1+\baru_2\cos\varphi}{1-\baru_2\cos\varphi}\right)\left(\frac{1+\baru}{1-\baru}\right)=\frac{1+\baru_1\cos\vartheta}{1-\baru_1\cos\vartheta}, \label{eq:t-ratio}
\end{equation}
where the second equality results from the velocity addition formula.
 If light is being used as the messenger, $\baru_1=\baru_2=1$. From the definition of the cross-ratio, \eqref{eq:x-ratio-bis}, which expresses distances as ratios instead of differences, the hyperbolic length of an arc subtended by angles $\varphi$ and $\vartheta$, with $\varphi>\vartheta$ is~\cite{mesch}
\begin{equation}
r/r_0=\half\ln\frac{(1+\cos\vartheta)(1-\cos\varphi)}{(1-\cos\vartheta)(1+\cos\varphi)}. \label{eq:hyp}
\end{equation}

The second formula of Lobachevsky relates the exponential distance between two horocycles, $\varrho$, to the hyperbolic cosine of the $h$-distance, $r$, viz.,
\begin{equation}
e^{\varrho/\kappa}=\cosh\left(r/r_0\right)=1/\sin\vartheta, \label{eq:lob-ii}
\end{equation}
where $\kappa$ is an absolute constant.
Introducing \eqref{eq:lob-ii} into the red-shift formula \eqref{eq:doppler-bis} gives
\begin{equation}
\nu=e^{-\varrho/r_0}\nu_0. \label{eq:red}
\end{equation}
An exponential law for the longitudinal red-shift, \eqref{eq:red}, has  been proposed~\cite[\S 6.4]{pro}, but not for the transverse red-shift. The linear approximation
\begin{equation}
\frac{\Delta\nu}{\nu_0}\simeq\frac{\varrho}{\kappa} \label{eq:linear}
\end{equation}
is usually quoted in texts on cosmology because the radius of curvature $\kappa$ is very large compared to $\varrho$. In general, the ratio $\varrho/\kappa$ will be of the order of the ratio of the Schwarzschild radius to the radius of a star, again a very small quantity. Therefore, non-Euclidean space time should manifest itself on small scales of the order of the Schwarzschild radius since the larger $\kappa$ is, in comparison to $\varrho$, the less non-Euclidean character of the Lobachesvkian plane.

 \section{comparison to \lq general\rq\ relativity}
 
 Einstein's theory of relativity essentially consists of two principles~\cite[p. 233]{fock}: The unification  of space and time into a four dimensional space with an indefinite metric, and the relation of the curvature of the space to the presence of matter. Einstein also proposed  an \lq equivalence\rq\ principle between inertia and gravitational mass, or between acceleration and gravitation. The latter has been criticized by Fock~\cite[pp. 232-233]{fock}, and by the present writer~\cite{lavenda}. Gravitational considerations appear only the the specification of the absolute constant, which is related to the constant, negative curvature of the hyperbolic space. Said differently, the centrifugal term appears explicitly in the metric, while the gravitational potential does not.
 
 In the general case of non-uniform motion the relevant space is the Friedmann-Lobachevsky velocity space~\cite[\S 94]{fock}, which we have derived here without any appeal to Einstein's equations and the unphysical assumption that matter must be \lq dust like\rq\ at zero pressure. The velocity components are related to the sides of a Lambert quadrilateral whose Weierstrass coordinates of the point of the acute angle show that the geometric mean time enters as a magnification of these coordinates, and not as a separate entity. Most importantly, by avoiding the \lq rigid scaffolding\rq\ employed by Einstein, which is applicable to inertial frames of reference only~\cite[p. 394]{fock},  acceleration has been accounted for as changes in velocity space, where the independence of the \lq coordinates\rq\ and time has disappeared.


\begin{thebibliography}{99}
\bibitem{greenberg}M. J. Greenberg, \emph{Euclidean and Non-Euclidean Geometries: Development and History\/}, 3rd ed. (W.H. Freeman and co., New York, 1993), p. 390.
\bibitem{busemann}H. Busemann and P. J. Kelly, \emph{Projective Geometry and Projective Metrics\/} (Dover, New York, 2006), p. 175.
\bibitem{lavenda}B. H. Lavenda, \emph{Hyperbolic nature of uniformly rotating systems and their relation to gravity\/}, arXiv:0804:1674.
\bibitem{gjw}G. J. Whithrow, \emph{Q. J. Math.\/} (Oxford) {\bf{4}}, 161-72 (1933).
\bibitem{bondi}H. Bondi, \emph{Cosmology\/} (Cambridge U. P., London, 1960).
\bibitem{milne}E. A. Milne, \emph{Kinematical Relativity\/} (Oxford U. P., London, 1948).
\bibitem{born}M. Born, \emph{Experiment and Theory in Physics\/} (Cambridge U. P., London, 1943).
\bibitem{moller}C. M\o ller, \emph{Theory of Relativity\/}  (Oxford U. P., London, 1952), p. 49.
\bibitem{whithrow}G. J. Whithrow, \emph{The Natural Philosophy of Time}, 2nd ed. (Clarendon Press, Oxford, 1980), p. 263.
\bibitem{page}L. Page, \lq\lq A new relativity,\rq\rq\ \emph{Phys. Rev.\/} {\bf 49}, 254-268 (1936).
\bibitem{pro}S. J. Prohovnik, \emph{The Logic of Special Relativity\/} (Cambridge U. P., London, 1967), p. 48.
\bibitem{fock}V. Fock, \emph{The Theory of Space, Time and Gravitation\/}, 2nd ed. (Pergamon Press, Oxford, 1969), p. 381 eqn (94.44) and eqn (94.48).
\bibitem{fred}F. Hoyle, G. Burbidge, and J. V. Narlikar, \emph{A Different Approach to Cosmology\/} (Cambridge U. P., Cambridge, 2000), p. 31.

\bibitem{mesch}H. Meschkowski, \emph{Noneuclidean Geometry\/} (Academic, New York, 1964), p. 78.







\end{thebibliography}
\end{document}